# Effect of complex orography on numerical simulations of a downburst event in Spain

**J. Díaz-Fernández[1,2*], C. Calvo-Sancho[1], P. Bolgiani[2], M. Sastre[2], M. López-Reyes[2,3], S. Fernández-González[4], M.L. Martín[1,5]**

[1] Department of Applied Mathematics. Faculty of Computer Engineering, Universidad de Valladolid, Spain.

[2] Department of Earth Physics and Astrophysics. Faculty of Physics, Universidad Complutense de Madrid, Spain.

[3]Astronomy and Meteorology Institute, Physics Department, University of Guadalajara, Guadalajara, Mexico.

[4] Agencia Estatal de Meteorología (AEMET). Spain.

[5] Interdisciplinary Mathematics Institute. Universidad Complutense de Madrid, Spain.

* Corresponding Author: Javier Díaz-Fernández (javidi04@ucm.es)

**Abstract**

A downburst is a localised and intense downdraft of air that descends quickly from the middle troposphere and reaches the Earth's surface. It is frequently originated by a thunderstorm or a supercell. Downburst winds can cause significant damage to buildings, infrastructure, and pose a great threat to aviation traffic. On July 1, 2018, many supercells were spotted near the Zaragoza Airport (Spain), and at least one of them generated a downburst that affected the airport, causing significant damage in the surrounding area. This event is here simulated using the Weather Research and Forecasting (WRF-ARW) numerical weather prediction model. Three different WRF-ARW orography experiments are carried out to investigate if the region's complex orography has an important role in supercell and downburst development over the research area. One of the three experiments uses the default orography as control; another one uses a 90% smoothed orography, and the third experiment is configured with a high-resolution dataset. Several atmospheric and convective variables are compared for each orography experiment. Results show that MUCAPE is clearly higher when the orography complexity is reduced. The smoothing process leads to a more uniform wind flow, contributing to the formation of numerous supercells. However, supercells channel through valleys and mountains in the control and high-resolution orography experiments, where the surface wind divergences are uniquely reproduced, and the highest reflectivity values are observed. Moisture advection from the Mediterranean Sea is essential in the process, reaching more deeply into the study region in the smoothed orography experiment due to the lack of orographic barriers. Orography affects dynamic and thermodynamic features, which have considerable effects on the formation and development of downbursts.

## 1. Introduction

Supercells are one of the atmospheric features that have a great influence on human activities due to their rapid development and the associated severe weather phenomena (large hail, strong winds or flash floods...), which can have destructive consequences (Fujita, 1980; Lee et al., 201; Taszarek et al., 2020a; Martín et al., 2021). A supercell is a convective storm that develops in a significantly vertically sheared environment and presents a deep and persistent mesocyclone. According to Browning (1962) and Quirantes (2008) it is the most organised, severe, and long-lasting form of mesoscale isolated deep convection phenomena. The formation of a supercell may lead in severe and extremely difficult phenomena to forecast (e.g., downbursts, tornadoes, and giant hail). Particularly, downbursts represent a major threat to air safety (Cao, 2020).

Fujita (1976) and Fujita and Byers (1977) first described atmospheric downbursts as intense downdrafts near the ground, causing surface damage and becoming a risk to aircraft at low altitudes. According to Fujita and Byers (1977) and Fujita and Wakimoto (1981), a downburst occurs when a strong downdraft in the lower troposphere produces a destructive surface wind with a divergent pattern, that extends from less than one to tens of km; these surface winds produce a toroidal-shaped gust front known as outflow. Fujita (1980) classifies downbursts according to the planetary horizontal scale of the outflow and the duration of the peak winds. A macroburst is defined by an outflow diameter greater than 4 km and a lifespan of more than 15 min, whereas anything below those limits is defined as a microburst. Moreover, downbursts may be classified into two types (Wakimoto, 2001): low reflectivity (dry) microbursts, with precipitation below 0.25 mm and radar reflectivity below 35 dBZ, and high reflectivity (wet) microbursts, with precipitation and reflectivity above the aforementioned limits. Microbursts represent a serious threat to the security of aircraft due to their small size and short duration (Fujita, 1980, 1985, 1990; Wolfson et al., 1994).

On July 1, 2018, a downburst was observed at Zaragoza Airport (Spain) at 16:50 UTC, with a maximum wind gust of 135 km/h (maximum gust record for the station), 24 mm of precipitation with intensities of 36 mm/10 min, and a temperature drop of 10.7 $^{o}$C during this severe convective weather phenomenon (see Figures 96-99 in Sanambrosio et al., 2019). This downburst was triggered by an anticyclonic cell formed by a storm splitting process (Sanambrosio et al., 2019).

High-performance computers provide high-resolution numerical models and simulations for studying downbursts, making these models an effective tool for understanding the relevant physical processes involved in downburst generation. Van Dijke et al. (2011) studied a downburst event coupled with a bow-echo structure, analysing the simulated downburst's winds and reflectivity obtained from the high-resolution Adavanced Research Weather Research and Forecasting model (WRF-ARW; Skamarock and Klemp, 2008). Bolgiani et al. (2020) simulated several episodes of high-reflectivity downbursts using the WRF-ARW model at spatial resolutions of 400 m; their results show a good performance of the model in accordance with observations.

The objective of this study is analysing the Zaragoza Airport downburst considering simulations obtained from the WRF-ARW model with three different schemes of orography: the standard WRF-AWR orography, a high-resolution dataset and a 90% smoothed orography. Several research have investigated the effects of orography on convective phenomena (Blumen, 1992; Gentile et al., 2014; Matsangouras et al; 2014; Leon-Cruz et al., 2019), sea breeze (Miao et al., 2003) and precipitation (Sotillo et al., 2003; Rotunno and Houze., 2007; Richard et al., 2007). However, to the authors' knowledge no research has been performed to study the importance of orography in the formation of a downburst event in Spain. This research aims to evaluate the WRF-ARW model's ability to reproduce downbursts and assess the influence of orography in the simulations.

The work is organised as follows: Section 2 describes the data and methodology used to produce the results, which are displayed and discussed in Section 3. Section 4 provides the study's conclusions.

## 2. Data and methodology

The following subsections describe the numerical weather prediction model configuration employed to simulate the three orography experiments associated with severe convective weather event that occurred at Zaragoza Airport on July 1, 2018, and the methodology description.

### 2.1. Experimental setup

To simulate the above-mentioned severe event, the WRF-ARW model (version 4.0.3; Skamarock et al. 2019) is used. The simulations are conducted with three domains centred over Zaragoza Airport (Figure 1a), under a two-way nesting strategy for a period of 24 h, allowing a 6 h spin-up and an initialization time at 00:00 UTC. Details about model configuration can be found in Table 1. In particular, the physics parametrizations here used are in accordance with related studies to severe convective weather phenomena in Iberia (Calvo-Sancho et al., 2020; Granda-Maestre et al., 2021) or microphysics sensitivity to extreme precipitation events (Tan, 2016; Eltahan and Magooda, 2018).

**Table 1.** Model configuration summary.

| | |
|---|---|
| **Initial/Boundary Conditions** | ERA5 reanalysis (0.25° horizontal resolution, 37 pressure levels, 1-hour temporal resolution) |
| **Number of Domains** | 3 (Two-way nesting) |
| **Horizontal resolution** | 15 km (D01), 3km (D02) and 600 m (D03) |
| **Vertical Levels** | 65 sigma pressure levels (first level at 910 hPa; higher resolution in lower troposphere for convective processes) |
| **Output Frequency** | D01, D02: every hour; D03: every 5 minutes |
| **Physics Parametrizations** | Shortwave/Longwave Radiation: New Goddard Scheme (Chou and Suarez, 1999; Chou et al., 2001)<br>Surface Layer: Revised MM5 (Jimenez et al., 2012)<br>Land Surface: 5-Layer (Dudhia, 1996)<br>Planetary Boundary Layer (PBL): BouLac (Bougeault and Lacarrere, 1989)<br>Microphysics: Aerosol-aware Thompson (Thompson and Eidhammer, 2014)<br>Cumulus Parametrization: Tiedtke scheme (D01) (Tiedtke, 1989; Zhang et al., 2011); Explicit (D02, D03) |
| **Orography Experiments** | CTRL: 30-second (~1 km) resolution GMTED terrain data (Danielson and Gesch, 2011)<br>NOORO: Reduced SRTM orography by 90% for smoother terrain (Miao et al., 2003)<br>HIRES: 90-meter resolution SRTM orography (Reuter et al., 2007) |

Three WRF-ARW experiments are generated to examine the complex orography that prevails in the region (Iberian System barrier and Ebro Valley) as a relevant factor in convective initiation, supercell storm formation and downburst development over Zaragoza Airport. The control (CTRL) experiment is set to 30-second (~1 km) horizontal resolution GMTED terrain data (WRF-ARW high-resolution default orography; Danielson and Gesch, 2011). According to Leon-Cruz et al. (2019) methodology, the non-orography simulation (NOORO) is set up to reduce the Shuttle Radar Terrain Mission (SRTM) orography by 90%, resulting in only 10% of the orography, in order to create a smoothed and plane effect just in D03 (Figure 1b). The smoothing of the orography by 90% implies a notable decrease in the ruggedness of the terrain in comparison to the original orography, as well as a reduction in elevation (Miao et al., 2003; Matsangouras et al., 2014; León-Cruz et al., 2019). This reduction preserves the general signal of the terrain, allowing us to visually discern geographical features and facilitate comparisons with other orography experiments. Finally, the high-resolution simulation (HIRES) runs with orography data from the 90-meter SRTM (Reuter et al., 2007). In the CTRL and HIRES experiments, the D03 maximum terrain height ranges from 2000 m to 200 m (Figures 1c, 1d). Therefore, the D03 average elevations are 853 m (HIRES), 852 m (CTRL), and 158 m (NOORO).

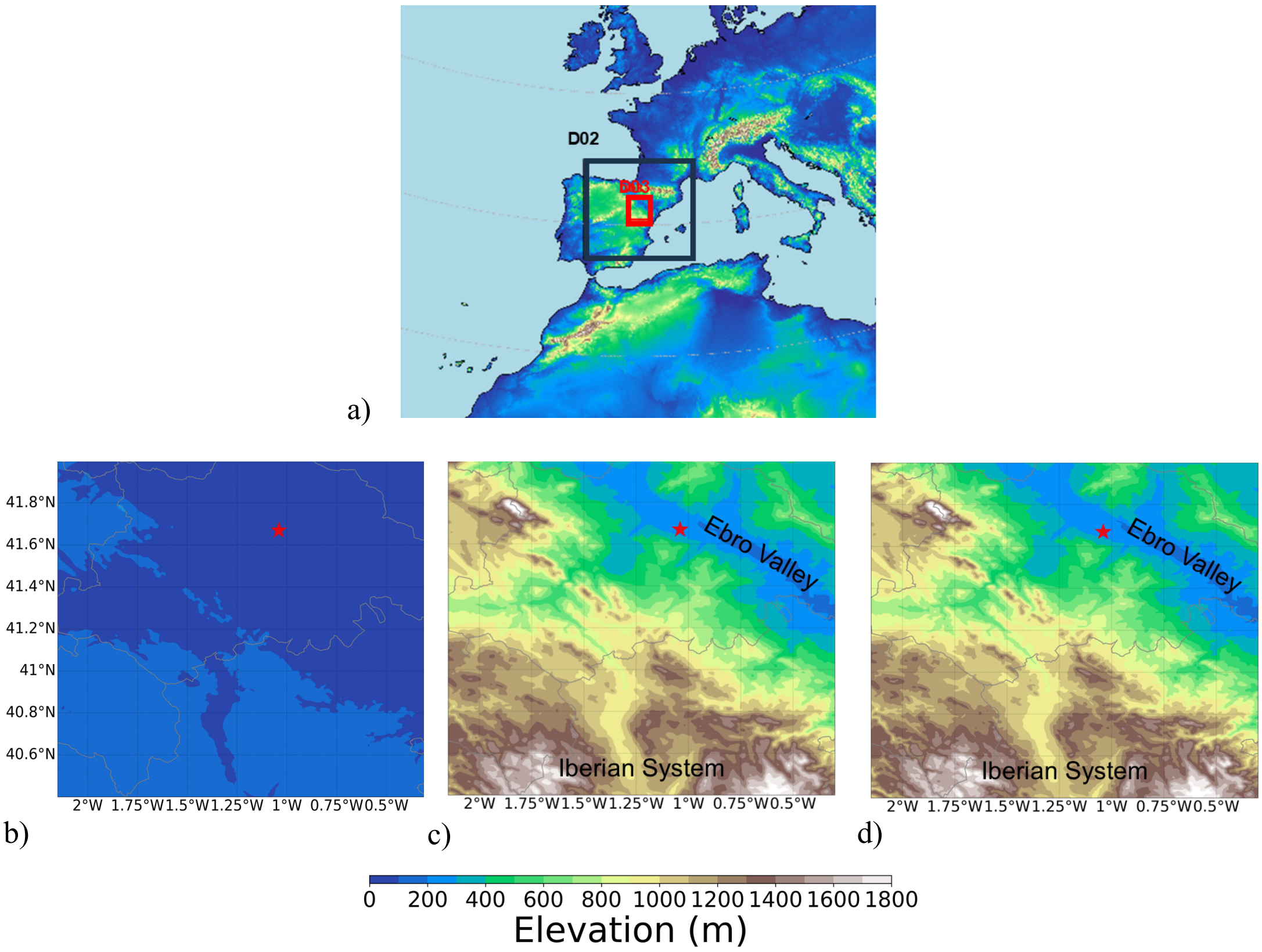

**Figure 1.** a) WRF-ARW domain configuration. Outer boundary corresponds to D01. Elevation maps of the windowed-in region of interest (d03) for the three different orography b) NOORO, c) CTRL and d) HIRES experiments. Red star indicates the Zaragoza Airport which is the location where observations come from.

### 2.2. Atmospheric and convective variables

To evaluate the pre-convective synoptic setting on the analysed day (July 1, 2018, at 14:00 UTC), the 500 hPa and 300 hPa geopotential height, the 2 m temperature and the mean sea level pressure fields from ERA5 reanalysis are used. From the WRF-AWR D03 simulations, several convective variables are calculated using the thundeR R language package (Taszarek et al., 2021) to characterise the convective environment of the downburst event and evaluate the similarities and differences between the different orography experiments. These variables are displayed at the time when the downburst simulation is most noticeable for each experiment: NOORO 15:00 UTC; CTRL 15:15 UTC; and HIRES 15:30 UTC.

Also, from D03 simulations, the Most Unstable Convective Available Potential Energy (MUCAPE) and the Convective Inhibition (CIN) are computed using virtual temperature adjustment (Doswell and Rasmussen, 1994). The Storm-Relative Helicity over 0–3 km (SRH03), the deep-layer bulk vertical Wind Shear over 0–6 km above ground level (WS06) and the Updraft Helicity (UH) are also estimated. These variables are chosen based on similar studies related to severe convective storms and are commonly used in operational forecasting (Thompson et al., 2003; Kaltenböck et al., 2009; Westermayer et al., 2017; Rodríguez and Bech, 2021; Davenport, 2021; Gensini et al., 2021; Calvo-Sancho et al., 2022; Romanic et al., 2022). Furthermore, the Downburst Environment Index (DEI), defined by Romanic et al. (2022), is also derived to evaluate climatological aspects of favourable downburst environments. Finally, a 5-km rectangular grid centred on the downburst location is considered to compute a representative mean sounding profile and the time series of the atmospheric and convective variables above mentioned as well as dewpoint, temperature at 2 m, cold pool strength and precipitable water are evaluated for each experiment at Zaragoza Airport. According to Markowski and Richardson (2011) and Parker (2014), using a composite sounding rather than a single point sounding is advantageous to study supercells or downbursts because it provides a more comprehensive understanding of atmospheric conditions and environment. A composite sounding allows to capture the spatial variability of several key variables, providing a more accurate representation of the atmospheric environment.

## 3. Results and discussion

### 3.1. Synoptic features

In this section the synoptic pattern of the downburst day is first presented to show the atmospheric conditions during the severe event genesis. The prevailing atmospheric conditions during the downburst event show a deep low, a so called cut off low, centred on the northwest coast of Iberia with a cold core of -16 ºC at 500 hPa (Figure 2a). This cut off low generates a cyclonic circulation throughout Iberia favouring mesoscale updrafts over the downburst location (Barnes et al., 2021; Huang et al., 2022). The typical pressure low due to thermal conditions over the downburst area, is shown in Figure 2b. These lows are usual during the summer, which favours moist flows from the Mediterranean Sea over the location of the event (Tullot, 2000; Calvo-Sancho et al., 2022). The upper and low atmospheric pattern configuration also favours surface convergence with southwesterly flows from the cut off low (Sanambrosio et al., 2019).

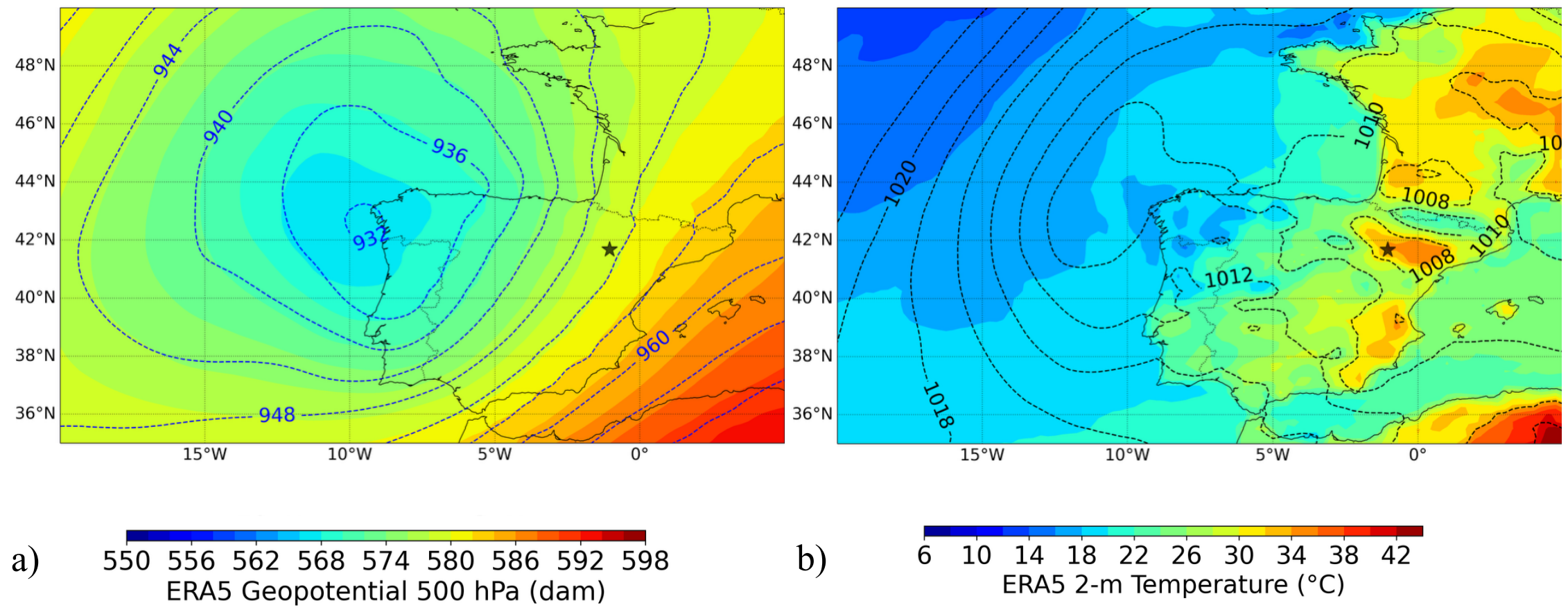

**Figure 2.** Atmospheric pattern configuration ERA5 synoptic charts for 2018-July-01 14:00 UTC: a) 500 hPa geopotential height (shaded; dam) and 300 hPa geopotential height (black contours; dam). b) 2 m temperature (shaded, ºC) and mean sea level pressure (black contours, hPa).

### 3.2. Dynamic and thermodynamic variables

The importance of orography in some of the atmospheric and convective variables linked to the formation of the convective weather phenomenon that happened in Zaragoza Airport on July 1, 2018, is now examined using three different orography experiments.

Figure 3 depicts the reflectivity pattern, the wind direction and speed, and the vertical cross section in the downburst location for each of the three experiments. It is worth noting that the CTRL (15:15 UTC) and HIRES (15:30 UTC) simulated downbursts are located about 30 km northeast of Zaragoza Airport (Figures 3b-c), while the wind flow follows a southwest-northeast track. Both, location and wind flow, are according to observations (Sanambrosio et al., 2019). However, in the NOORO experiment (Figure 3a), the downburst (15:00 UTC) is produced by a large group of cells with high reflectivity located to the east of Zaragoza Airport, moving south to north. According to the observational report by Sanambrosio et al. (2019), the downburst that hit Zaragoza Airport at 16:50 UTC was triggered by an anticyclonic cell formed by a storm-splitting process, whose cyclonic cell weakened and then intensified again. It is significant that the higher orography resolution experiment simulates the downburst at a closer time (80 min) to the event, while the CTRL simulates the downburst 95 min before the observation report and the NOORO 110 min before. This is not uncommon for the model as the WRF-ARW simulation of supercells has been noted to begin earlier and move faster than observations as established in prior studies (Taszarek et al., 2019; Pilguj et al., 2019; Bolgiani et al., 2020). This may also be the reason for the different location and timing of the event.

The time reflectivity evolution from the CTRL and HIRES experiments (not shown) suggests how successfully the WRF-ARW model simulates the storm splitting process, with CTRL performing better, showing two different cells, being the westernmost a weakening cyclonic cell, as observed by Sanambrosio et al. (2019). According to these observations, the maximum reflectivity from radar data during the downburst event is 60 dBz, with echo tops reaching 14 km. However, the maximum reflectivity simulated here is somewhat lower, reaching 45–55 dBz in the surroundings of Zaragoza Airport. This reflectivity underestimation can also be found in another related studies. Lompar et al. (2017) use the WRF-ARW to simulate a severe convective storm in Serbia and find reflectivity values that were 10 dBz lower than radar data. Pilguj et al. (2019) detect a similar underestimation in an isolated tornadic supercell WRF-ARW simulation in Poland.

Evaluating wind direction and speed, it can be seen how the maximum wind speed in the domain matches the reflectivity distribution (Figures 3d-f). The three experiments simulate wind speeds of about 16–20 m/s in the downburst location, with the HIRES maximum speed values greater than 22 m/s. The reader should note that these are not gust wind values, but only the maximum wind speed, so it cannot be directly compared to the observed 38 m/s gust observation, although it gives us an approximation of the event intensity. The results show a wind divergence point corresponding to the strongest winds and highest reflectivity values, indicative of a potential downburst. Furthermore, the surface winds with the characteristic toroidal-shaped outflow associated with downburst (Fujita, 1980; Fujita and Wakimoto, 1981) is clearly depicted in HIRES (Figures 3f, i). However, the gust front shown in the NOORO is produced by the supercell set (Figure 3d).

Figures 3g-i depict vertical wind and reflectivity cross sections at the simulated downburst location and time. The cross sections show a 40–50 dBz descendent core reaching the surface in the three studies with a considerable downdraft. The HIRES experiment recreates the most realistic downburst, with the downdraft pulling down the reflectivity

core and a maximum vertical velocity of -11 m/s, which is lower for CTRL and NOORO (-7 m/s). Low level surface divergence is observed (most clearly in HIRES), with an updraft to the north of the downburst (most noticeably in CTRL) caused by interaction with prior winds. The HIRES (CTRL) cross section displays the 50 (40) dBZ core reaching the ground, and the outflow winds. On the other side, the NOORO experiment reveals a 40 dBz core hitting the ground. No surface divergence is noticeable since the downdraft does not reach the surface at the moment. It is necessary to note that the selected downburst times are not equivalent and due to the temporal output resolution (5 min), the precise time when the downburst reaches the surface is most likely not captured by some experiment. Bolgiani et al. (2020) find Comparable reflectivity values (about 50 dBz) and vertical wind speed (-6 to -8 m/s) were found in a WRF-ARW high-resolution simulation of a series of microbursts observed in the USA (Bolgiani et al; 2020) and a large hailstorm event on the Adriatic Sea (Ricchi et al., 2023). These authors also detect wind divergence on the surface. In the current study, we can clearly assume that the NOORO, CTRL, and HIRES experiments reproduce the downbursts, each with its own peculiarities as stated above.

The three experiments show considerable differences in the location of maximum wind speed and reflectivity (Figures 3a – f). One of the key reasons for this is the variation in orography across the three experiments, leading to different simulated pressure and temperature gradients that may explain the different locations of the wind maxima (León-Cruz et al., 2019). These orographic variations are most evident in the NOORO experiment, where elevation differences of up to 1200–1400 m can be observed in the southern half of the study domain compared to the CTRL and HIRES experiments (Figure 1). On the other hand, the fact that the model simulates a deep convection event while almost completely removing the orography suggests that the shown convective environment is powerful enough to generate free convection and conducive to the formation of supercells.

Smoothing the orography (NOORO) reduces terrain roughness, decreasing resistance to wind and resulting in a more constant flow. While the surface wind speeds increases when the roughness is reduced (Figure 3d), the absence of mountains prevents them from acting as a barrier to wind flow but also to trigger a mechanical uplift. In CTRL and HIRES, the results show changes in wind direction and speed because the lower air masses interact with the mountains located at the southern part of the domain (Figures 3e, f). As a result, the smoothed orography allows greater acceleration and higher wind speeds near Zaragoza Airport.

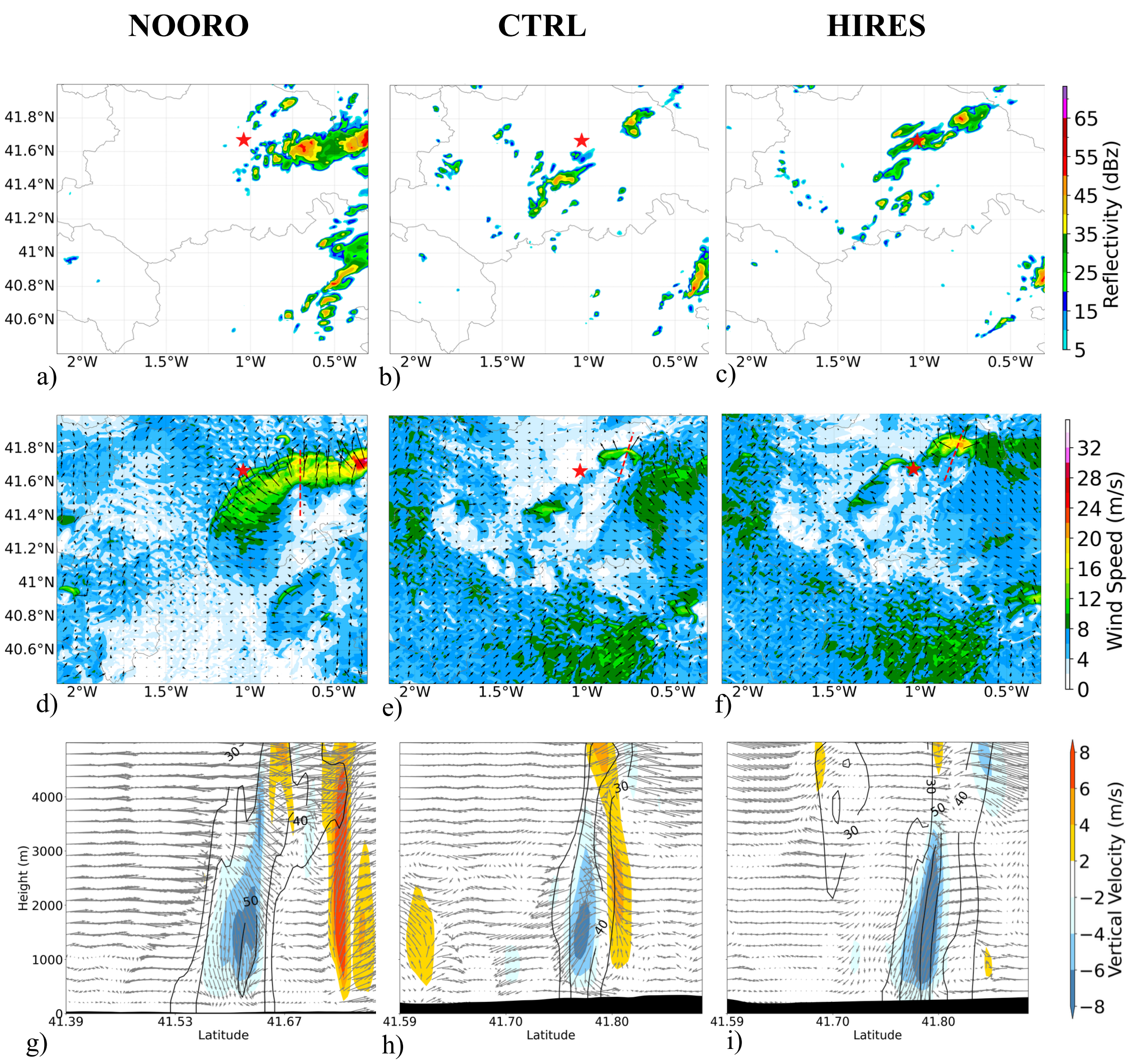

**Figure 3.** Reflectivity (dBz) distributions for a) NOORO, b) CTRL and c) HIRES; 10-m wind speed (shaded; m/s) and wind direction (arrows) for d) NOORO, e) CTRL and f) HIRES; Vertical cross section ( red line in Figures 3d-f) for vertical wind vectors (arrows) and speed (shaded) and reflectivity (contours) for g) NOORO, h) CTRL and i) HIRES, on July 1, 2018.

Concerning MUCAPE and CIN (Figure 4a-c), the NOORO experiment reveals the highest MUCAPE values (>1600 J/kg) in the domain since the moist air flow from the Mediterranean reaches the study area due to the absence of mountains in its path. The MUCAPE values in both the CTRL and HIRES simulations are comparable and lower than NOORO, with the lowest MUCAPE values located in high terrain areas. These results are consistent with those of León-Cruz et al. (2019), who investigated the role of complex terrain in the generation of tornadoes in Mexico, yielding higher MUCAPE values in smoothed topography experiments. On the other hand, CIN values smaller than 100 J/Kg can be found near the downburst locations for the three experiments; however, these values do not present a significant challenge to the development of convection (Quirantes et al., 2014; Calvo-Sancho et al., 2022) because of the high MUCAPE values.

To explain the differences in the unstable environment among the three experiments, significant variations in the virtual temperature were evaluated (not shown). In the NOORO, the moisture flow from the Mediterranean Sea reaches the study area without any orographic barriers, resulting in higher virtual temperature values. Also, the smoothed orography in NOORO allows warm moist air to flow into the region and lower the Level of Free Convection (LFC), enhancing instability. In contrast, the moist flow is channelled by the terrain and the LFC is higher in the HIRES and CTRL experiments. The presence of orography in these experiments acts as a barrier that modifies the moisture flow and vertical temperature profile, thus influencing the MUCAPE and CIN values. Therefore, orography seems to be an important factor, conditioning and already unstable environment by altering the distribution of moisture and virtual temperature profiles.

WS06 and the SRH03 distributions for the three simulations are displayed in Figures 4d–f. The intensity of WS06 in the downburst area are relatively similar (values less than 8 m/s) in the three experiments, since once the supercell has passed through, changes in wind direction and speed reduce and energy dissipation increases, resulting in weaker WS06. However, the WS06 values located right north of the downburst (i.e., the WS06 values that the downburst could have found if it followed the same trajectory) range from 26 to 22 m/s, with the NOORO experiment revealing a slightly higher value. These results are consistent with previous investigations of supercells in the Iberian Peninsula (Calvo-Sancho et al., 2022) and the simulated sounding at Zaragoza Airport using the HARMONIE-AROME model (Sanambrosio et al., 2019), where a WS06 of 20 m/s is obtained. Related to SRH03, Calvo-Sancho et al. (2022) suggest that the SRH03 threshold for supercell development in Spain is 100 $m^2/s^2$. Figures 4d-f show that in the surroundings of the downburst area, SRH03 values are around 200 $m^2/s^2$. These results, together with MUCAPE values higher than 800 J/kg, indicate that rotation is highly plausible in the convective cells. It is also interesting to note that the maximum SRH03 (500 $m^2/s^2$) is located at the southeastern domain in the NOORO experiment.

Finally, Figures 4g-i depict the 2 m temperature and UH. The most noticeable feature of these results is the appearance of a "temperature arc" southeast of the Zaragoza Airport in the NOORO experiment (Figure 4g), which is clearly matching the gust front created by the several cells simulated (Figure 3d). Furthermore, UH values larger than 100 $m^2/s^2$ are located northward of this arc, indicating potential supercell development zones. The CTRL and HIRES experiments lack the abovementioned extensive arc, however both show pools of lower temperatures (25–23 ºC for CTRL and 20–23 ºC for HIRES; Figures 4h–i) matching the position of the possible downbursts. The temperature is lower for the HIRES, in accordance with the stage at which the downburst is captured, as the vertical profile shows that the downburst for this simulation is fully developed and has clearly reached the ground (Figure 3i), while the CTRL downburst was captured at the initial moment of reaching the surface (Figure 3h). These patches of cooler temperatures indicate air descents from upper atmospheric levels and heavy rainfall, presenting a proper simulation of the downburst (Bolgiani et al., 2020). In the context of the NOORO experiment, temperatures in the vicinity of Zaragoza Airport range around 35 ºC. CTRL and HIRES experiments show values ranging from 30 to 33 ºC, respectively. Compared to the temperature observations (Sanambrosio et al., 2019), which was 31 ºC at the Zaragoza Airport, the CTRL experiment better simulates near-surface temperature.

The UH is an important parameter for forecasting supercell thunderstorms, related with vertical speed of upward and relative vorticity. The UH informs about areas where rotating updrafts are expected (Taszarek et al., 2019; Pilguj et al., 2019). In the current study, UH values greater than 100 $m^2/s^2$ are depicted to indicate the cell's most active regions. This threshold is significant because it highlights areas with strong rotational updrafts, suggesting a higher likelihood of severe weather phenomena. While the NOORO experiment (Figure 4g) shows large UH spatial values to the north-northwest of the domain, the CTRL (Figure 4h) and HIRES (Figure 4i) experiments depict high UH values just located south of Zaragoza Airport, corresponding to the cooler temperature areas and potential supercells (Figure 3b,c). These differences in UH spatial distribution between the three experiments can be explained by the influence of orography on atmospheric flow. In the NOORO, UH distributes along to the gust front (where the reflectivity is maxima) generated by the absence of orography. However, in the CTRL and HIRES, the presence of complex terrain can modify local wind patterns, leading to enhanced convergence and localized vorticity.

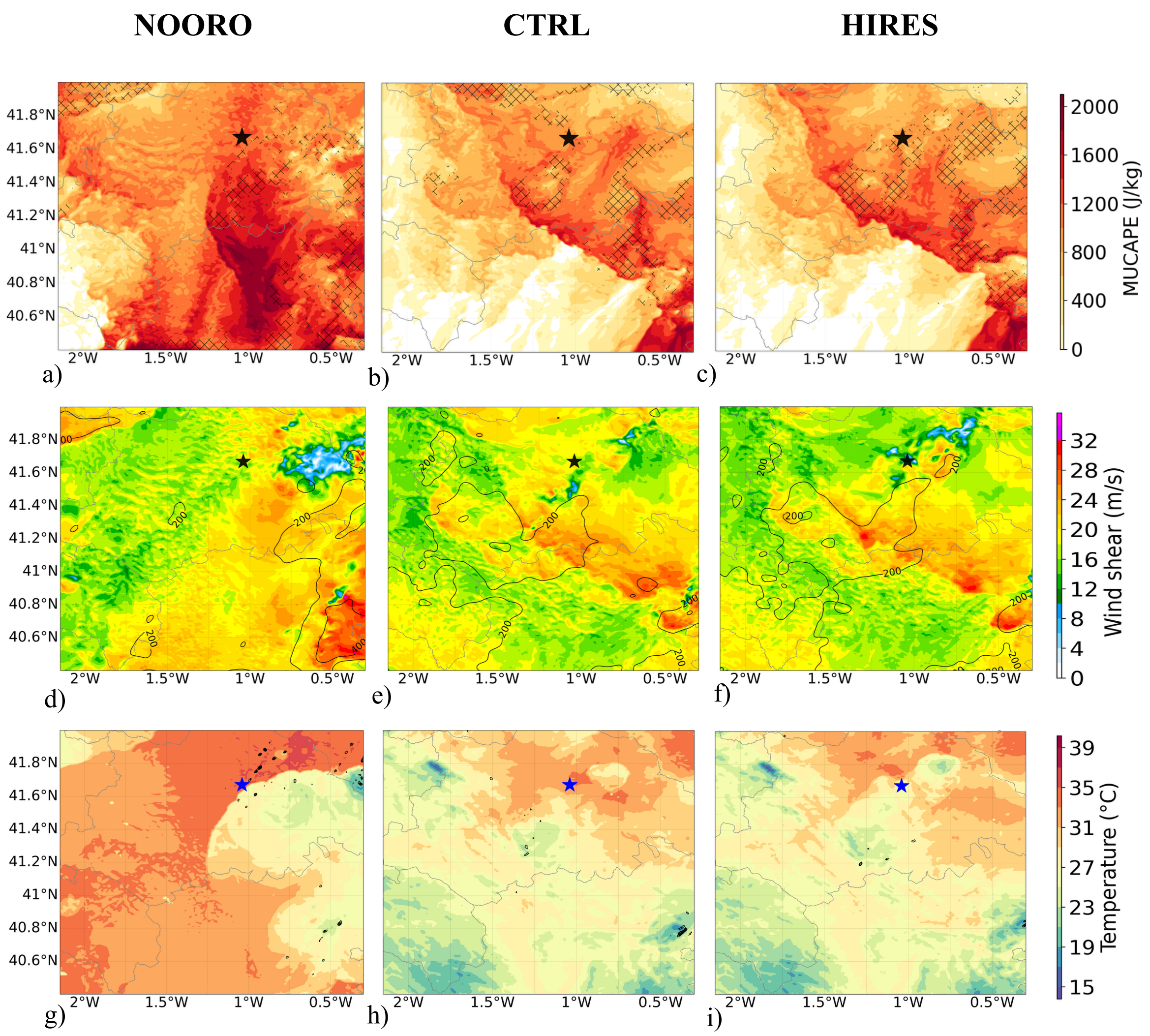

**Figure 4**. MUCAPE (shaded; J/kg) and CIN < 100 J/kg (cross) distributions for a) NOORO, b) CTRL and c) HIRES; WS06 (shaded; m/s) and SRH03 (contours; $m^2/s^2$) distributions for d) NOORO, e) CTRL and f) HIRES; 2 m temperature (shaded; J/kg) and UH (contours higher than 100 $m^2/s^2$) distributions for g) NOORO, h) CTRL and i) HIRES, on July 1, 2018.

Figure 5 depicts the simulated atmospheric soundings at downburst locations for the three experiments. In order to examine the closest preconvective environment, the simulated soundings are shown 15 - 20 min prior to the downburst time. At lower levels, the three simulations reveal profiles similar to the characteristic inverted V conductive to downbursts and high-reflectivity microbursts (Atkins and Wakimoto, 1990; Bolgiani et al., 2020). Furthermore, the experiments accurately reproduce the dry layer at middle levels indicating the possible entrainment of environmental dry air at midlevel layers, which can enhance downdrafts by promoting evaporation (Bolgiani et al. 2020). Downbursts generally begin at midlevel layers just above the 0 ºC level (Proctor, 1989). In an idealised downburst, the inverted V vertex aligns with the 0 ºC isotherm. Here, the closest distance between the vertex and 0 ºC isotherm corresponds to CTRL. It is also noteworthy that the NOORO sounding has an additional value of 250 m of super adiabatic environment to accelerate the downburst due to the lower elevations of the terrain.

Surface temperatures range between 33-34 ºC for the three experiments. MUCAPE values corresponds to 1189 J/kg for NOORO and around 750 J/kg for CTRL and HIRES. As expected, the lowest Lifting Condensation Level (LCL) belongs to NOORO. As in previous results, wind speed and direction show minimal variability for NOORO with strong south-southwesterlies winds along the whole vertical profile; the wind speed mean in the first 6 km is 11.3 m/s, being 9 m/s for the CTRL and HIRES experiments. The NOORO sounding indicates a deep wet layer (from 2.5 to 4 km), while in the remaining experiments it is shallower (from 3 to 3.5 km); above these wet layers, the dry layer is identified. In terms of the lifted index, the three studies demonstrate overall instability with values less than -2 ºC, showing the NOORO experiment the lowest value (-5 ºC).

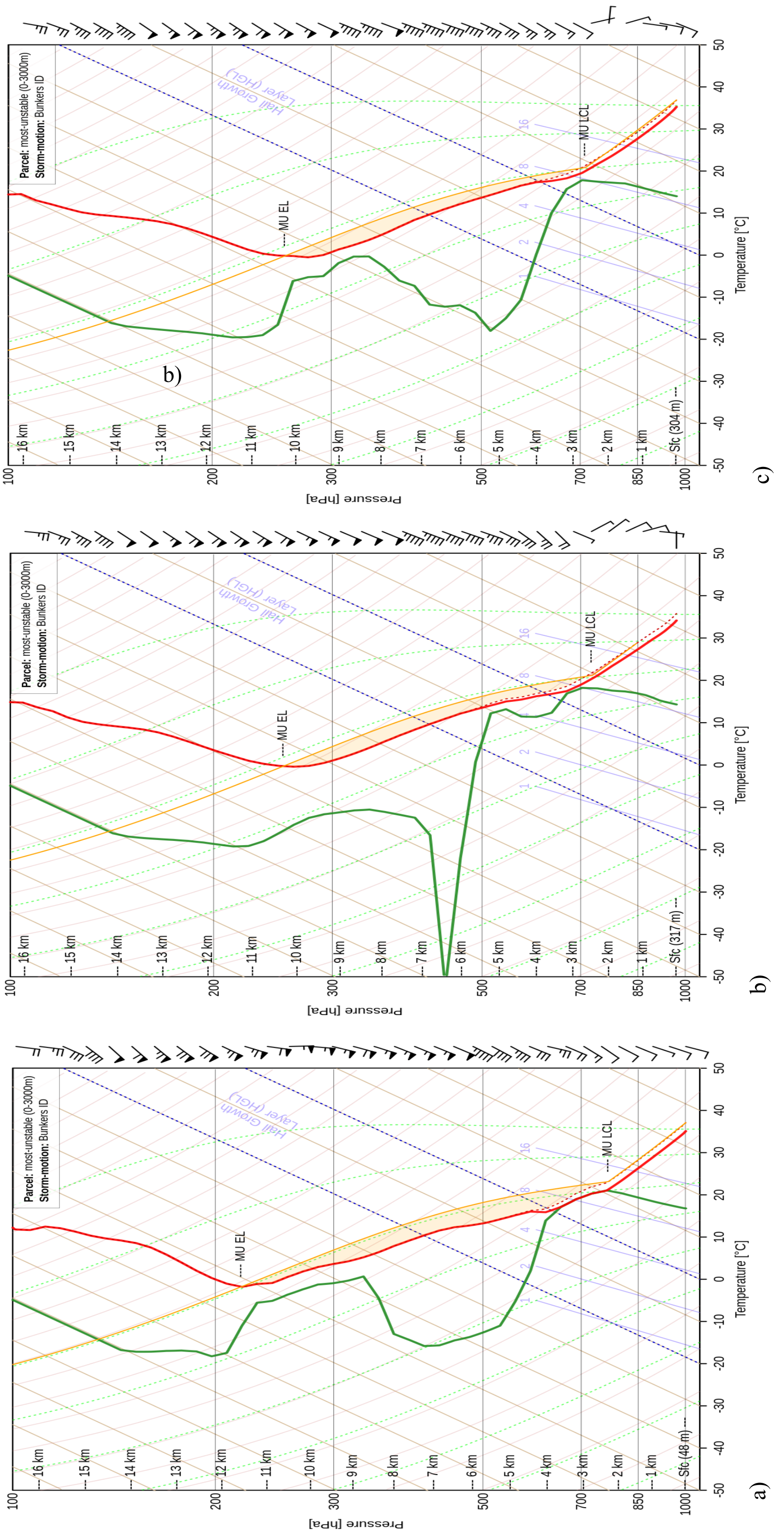

**Figure 5.** WRF-ARW soundings on July 1, 2018, in the downburst location 15 – 20 min before the downburst time for a) NOORO, b) CTRL and c) HIRES.

### 3.3. Time series

Concerning the time series of the atmospheric and convective variables, the surface temperature is first evaluated. The behaviour of the variable is as expected in a downburst event, showing a sharp decrease as the downburst core reaches the ground. The NOORO experiment presents a higher 2m temperature (Figure 6a), as expected by the lower elevation. Moreover, the temperature falls by around 8 ºC during the downburst event in the HIRES and NOORO experiments, whereas the CTRL has a smaller drop (4 ºC). The behaviour of the 2m dewpoint (Figure 6b) is the opposite to that of the 2 m temperature, increasing as the downburst reaches the ground, with the NOORO simulation showing the maximum values while the remaining experiments show values 2-3 ºC lower. It is interesting to note that the NOORO simulations presents a second peak for both variables, indicating that the downburst was not produced by an isolated cell as seen in Figure 3d. The wind speed evolution (Figure 6c) displays a maximum wind speed for HIRES (36 m/s), which is much lower for the CTRL (23 m/s) and NOORO (20 m/s) experiments at the downburst time (Figures 3d-f). In any case, all of the values are consistent with a downburst, and the differences are in line with the different stages at which each simulation captured the event (Figures 3g-i).

In the MUCAPE time series (Figure 6d), orography seems to play an essential role, resulting in a moderately unstable environment around the downburst time (between 700 and 1000 J/kg), with the exception of the NOORO experiment which shows severely unstable MUCAPE values (between 1250 and 1500 J/kg). This is in line with Leon-Cruz et al. (2019) who found higher MUCAPE values in smoothed orography tests in tornado investigations with several orography tests. The values are also consistent with Calvo-Sancho et al. (2022) who established threshold MUCAPE values ranging from 700 to 1800 J/kg to assess the occurrence of supercells in Spain. Concerning the CIN results (Figure 6e), the three experiments reproduce an unstable preconvective environment with values near to 0 J/kg prior to the downburst time and a significant drop (down to -100 J/kg) after the event. Figure 5f shows that the three experiments have moderate WS06 values, reaching 24 m/s for CTRL, 22 m/s for NOORO, and 18 m/s for HIRES, just prior to the downburst.

The cold pool strength is defined as the difference between the ambient temperature and the downdraft at the surface and, according to Romanic et al. (2022), it is recognised as the most skillful parameter for downburst identification. As it increases, the likelihood of a downburst improves. The cold pool strength time series (Figure 6g) reveals values around 16 ºC for HIRES and NOORO experiments and 14 ºC for CTRL. The obtained values are between the 75th and 90th percentile of ERA5 convective environments leading to downburst occurrences in the United States (Romanic et al., 2022).

In line with the previous varibale, Romanic et al. (2022) developed the DEI index for studying favourable downburst environments. It combines the cold pool strength and WMAXSHEAR (a square root of two times MLCAPE multiplied by 0-6 km wind shear; Taszarek et al., 2020b). Values over 0 suggest a favourable downburst environment with a proportion of the most probable occurrence as the DEI increases. In the current study, preconvective DEI values are positive for all three experiments (Figure 6h), indicating a favourable downburst environment. It is noticeable that the NOORO simulation produces the highest DEI results (exceeding 2), due to the higher CAPE values than the remaining experiments. Finally, precipitable water, as a measure of the total amount of water vapour in an atmospheric column, shows very similar values (around 40 mm) for the three experiments (Figure 6i), despite the differences in the dewpoint (Figure 6b). This could be because, while the atmospheric column in the NOORO experiment is deeper due to lower terrain elevation, the absence of orographic lifting and moisture dispersion leads to less precipitable water. In the HIRES and CTRL studies, the presence of orography forces moist air to climb, cool, and condense, increasing moisture concentration and resulting in more precipitable water despite a shorter atmospheric column above the elevated terrain and a lower dew point at surface (Smith, 1979; Houze, 2012).

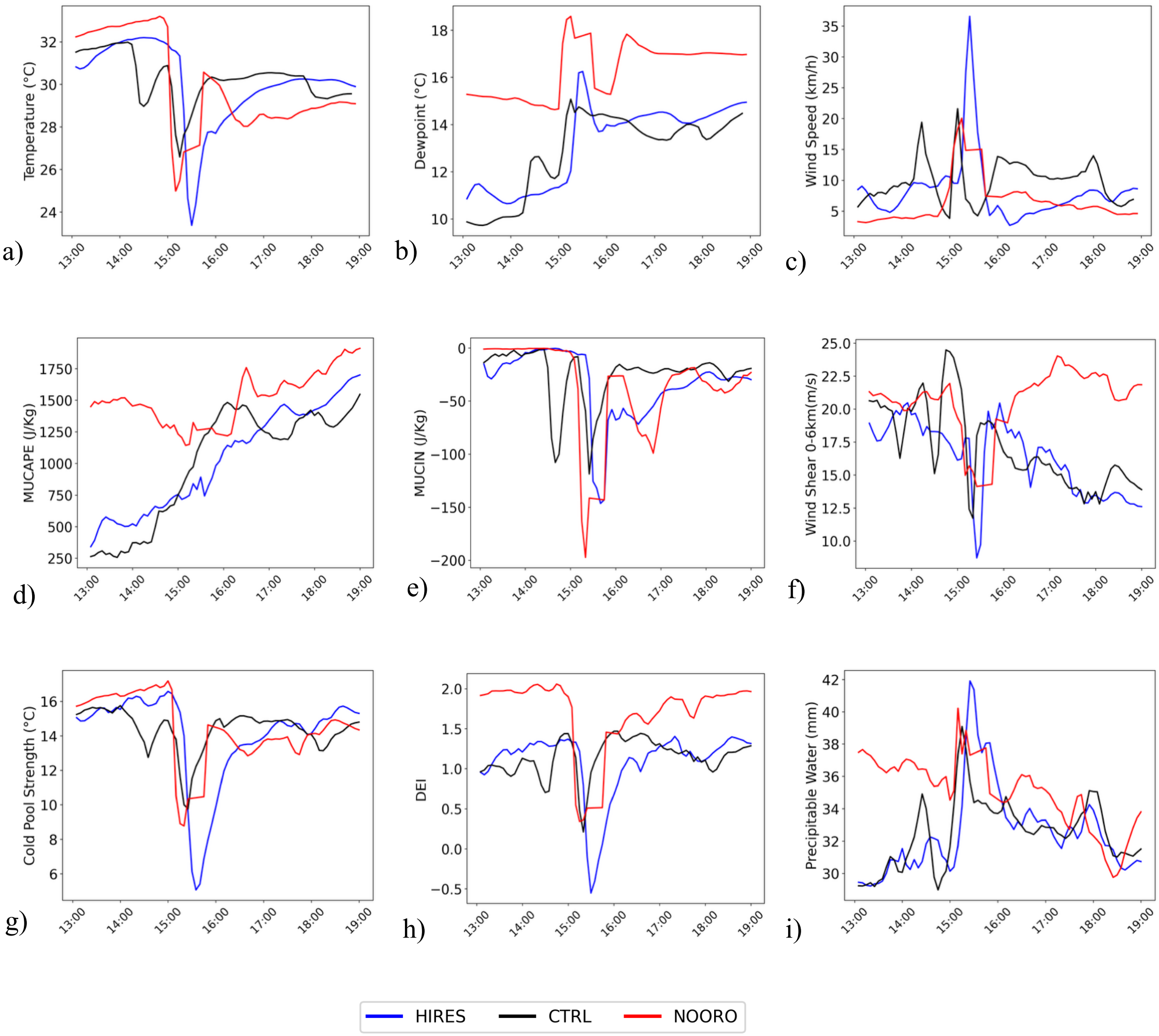

**Figure 5.** Time series of WRF-ARW (a) 2 m temperature (ºC), (b) 2 m dewpoint (ºC), (c) wind speed (m/s), (d) MUCAPE (J/kg), (e) MUCIN (J/kg), (f) WS06 (m/s), (g) Cold Pool Strength (ºC), (h) DEI and (i) Precipitable water (mm). Every variable is the average from at a 5 km grid to centre at the downburst location for NOORO (red line), CTRL (black line) and HIRES (blue line) experiments.

## 4. Summary and conclusions

In this article, the downburst event that occurred on July 1, 2018, at the Zaragoza Airport (Spain) is studied. The atmospheric environment of this event is reproduced in three experiments using the WRF-ARW model, in which the orography is modified to investigate how the orography influences the WRF-ARW's ability to reproduce the phenomena under consideration. The three experiments used are NOORO (90% smoothed orography), CTRL (default WRF-ARW orography) and HIRES (90-meter SRTM orography).

The downburst was observed at 16:50 UTC near Zaragoza Airport. However, in all three experiments, the WRF-ARW model simulated a downburst around 30 km northeast of Zaragoza Airport. Furthermore, it is worth noting that the greater the resolution of the orography, the closer the simulation is to the observation time of the downburst; the NOORO experiment reproduces the event at 15:00 UTC, CTRL at 15:15 UTC, and HIRES at 15:30. These differences in time and location are not uncommon in WRF-ARW for deep convection simulations.

The results indicate that the occurrence of this downburst event is associated with supercells activity. It is noteworthy that the synoptic scale analysis shows a suitable environment for the generation of supercells in the research area, with high MUCAPE values and humidity advection from the Mediterranean Sea, particularly notable in the NOORO experiment (>1600 J/kg) due to the absence of the orographic Iberian System barrier. Orography seems to be a key factor, providing an unstable environment modulated by the terrain's orography with lowest MUCAPE values in mountain areas. Orography also influences atmospheric instability and moisture distribution, enhancing localized convective activity. In the NOORO simulation, UH is more uniformly distributed, while the presence of complex terrain (CTRL and HIRES) enhances convergence and localized vorticity, leading to higher instability and potential for severe weather phenomena.

The WRF-ARW model successfully replicates the storm splitting process, with the CTRL performance better than the remainder experiments, indicating two different cells and a weakening cyclonic cell, in line with observations. The greatest reflectivity simulated in the downburst zone show values of 45–55 dBz for the three experiments, significantly lower than observed (60 dBz). CTRL and HIRES experiments simulated a 40-50 dBz core falling and reaching the surface, resulting in wind surface divergence. However, the downdraft in the NOORO doesn't reach the surface, most probably because of the output timing. Simulated atmospheric soundings from the three experiments in the downburst location show an inverted V pattern in the lower troposphere, suitable to downbursts and high-reflectivity microbursts.

The HIRES experiment simulates a characteristic toroidal-shaped outflow associated to the downburst. On the other hand, the smoothing of orography has been shown to lower wind resistance, allowing for faster acceleration and higher wind speeds due to the absence of wind-channelling over valleys and mountains, which typically occurs in locations with natural, unaltered terrain. The smoothed terrain produces a more continuous flow of wind that can contribute to the formation of many supercells and a gust front. In contrast, areas with a more realistic orography (CTRL and HIRES) generate storms channelled over valleys and mountains, resulting in fewer supercells and, at least in one case, a downburst. The influence of orography on downburst generation on July 1, 2018, can be related to the instability in the research region, enough to allow the development of deep convection. The HIRES experiment is the only one in which the downdraft (-11 m/s) makes the reflectivity core (50 dBz) to reach the surface developing wind divergences. Even though each of the three experiments produced DEI values greater than 1, NOORO reached a value of 2 during the downburst, suggesting a very favourable downburst environment. Cold pool strength peaks at 16 $^{o}$C for HIRES and NOORO, suggesting convective conditions leading to downbursts.

In conclusion, although the unstable preconvective environment is presented in such an event, the orography has played a significant influence on the distribution and intensity of several supercells in the studied area. Smoothed orography has promoted the development of many supercells by reducing wind resistance, whereas natural orography channels airflow, leading to the formation of fewer, but potentially more intense, convective systems. As a result, in the Zaragoza Airport downburst, dynamic and thermodynamic parameters are influenced by orography having an important effect on its genesis and development.

**Acknowledgements**

Javier Díaz-Fernández thanks the Spanish Ministerio de Ciencia, Innovación y Universidades for granting a Margarita Salas contract from Complutense University of Madrid using Next Generation funding from the EU. Carlos Calvo-Sancho acknowledges the grant support from the Spanish Ministerio de Ciencia, Innovación, y Universidades (FPI programmes PRE2020-092343). Moreover, this research has been supported by the Ministerio de Ciencia e Innovación project PID2023-146344OB-I00 (CONSCIENCE) and the ECMWF Special Projects (SPESMART and SPESVALE).

**Conflicts of Interest**

The authors declare no conflict of interest. The funding sponsors have no participation in the execution of the experiment, the decision to publish the results, nor the writing of the manuscript.